\documentclass{aa}

\usepackage{graphicx}
\usepackage{txfonts}
\DeclareRobustCommand{\VAN}[3]{#2}
\let\VANthebibliography\thebibliography
\def\thebibliography{\DeclareRobustCommand{\VAN}[3]{##3}\VANthebibliography}

\usepackage{color}

\usepackage{textcomp}
\usepackage{amsmath}	
\usepackage{amssymb}	
\begin{document}

   \title{The first high-redshift cavity power measurements of cool-core galaxy clusters with the International LOFAR Telescope}
   \titlerunning{The first high-redshift cavity power measurements of cool-core galaxy clusters with the ILT}

   \author{R. Timmerman\inst{1,2}\fnmsep\thanks{E-mail: roland.timmerman@durham.ac.uk (RT)}
        \and
        R. J. van Weeren\inst{1}
        \and
        A. Botteon\inst{3}
        \and
        H. J. A. Röttgering\inst{1}
        \and
        L. K. Morabito\inst{2,4}
        \and
        F. Sweijen\inst{1}
    }

   \institute{Leiden Observatory, Leiden University, P.O. Box 9513, 2300 RA Leiden, The Netherlands
              \and
              Centre for Extragalactic Astronomy, Department of Physics, Durham University, Durham DH1 3LE, UK
              \and
              INAF - Instituto di Radioastronomia, via P. Gobetti 101, 40129 Bologna, Italy
              \and
              Institute for Computational Cosmology, Department of Physics, University of Durham, South Road, Durham DH1 3LE, UK
    }

   \date{Received XXX; accepted YYY}

 
  \abstract{Radio-mode feedback associated with the active galactic nuclei (AGN) at the cores of galaxy clusters injects large amount of energy into the intracluster medium (ICM), offsetting radiative losses through X-ray emission. This mechanism prevents the ICM from rapidly cooling down and fueling extreme starburst activity as it accretes onto the central galaxies, and is therefore a key ingredient in the evolution of galaxy clusters. However, the influence and mode of feedback at high redshifts (\(z\sim1\)) remains largely unknown. Low-frequency sub-arcsecond resolution radio observations taken with the International LOFAR Telescope have demonstrated their ability to assist X-ray observations with constraining the energy output from the AGNs (or "cavity power") in galaxy clusters, thereby enabling research at higher redshifts than before. In this pilot project, we test this hybrid method on a high redshift (\(0.6<z<1.3\)) sample of 13 galaxy clusters for the first time with the aim of verifying the performance of this method at these redshifts and providing the first estimates of the cavity power associated with the central AGN for a sample of distant clusters. We were able to detect clear radio lobes in three out of thirteen galaxy clusters at redshifts \(0.7<z<0.9\), and use these detections in combination with ICM pressures surrounding the radio lobes obtained from standard profiles to calculate the corresponding cavity powers of the AGNs. By combining our results with the literature, the current data appear to suggest that the average cavity power peaked at a redshift of \(z\sim0.4\) and slowly decreases toward higher redshifts. However, we require more and tighter constraints on the cavity volume and a better understanding of our observational systematics to confirm any deviation of the cavity power trend from a constant level.
}


   \maketitle
%

\section{Introduction}

\begin{table*}
    \caption{Summary of the sample of galaxy clusters used in this paper. High-resolution LOFAR observations have been processed for the galaxy clusters indicated with boldface. The total flux density of each source is listed as reported by three radio surveys: TGSS at 150~MHz, VLASS at 3~GHz, and LoTSS at 144~MHz. Non-detections are indicated with a horizontal dash. Sources which are located beyond the coverage of the corresponding survey are indicated with ellipsis.}
    \centering\small
    \begin{tabular}{lllllllll}\hline\hline
        Cluster name & R.A. (J2000) & Dec. (J2000) & \(z\) & \(c_\mathrm{SB}\)\ & M\(_{500}\) & S\(_\mathrm{TGSS}\) & S\(_\mathrm{VLASS}\) & S\(_\mathrm{LoTSS}\) \\
                    &               &              &       &                    & (\(10^{14}\ \mathrm{M}_\odot\)) & (mJy) & (mJy) & (mJy)\\\hline
        RCS\,1419+5326                & 14h19m12.1s & 53d26m11.6s & 0.620 & 0.185 & 2.60 (2) & -    & -    & -\\
        CDGS40                        & 14h50m09.0s & 09d04m48.7s & 0.644 & 0.103 & 1.86 (4) & -    & -    & ...\\
        EGSXG\,J1417.9+5235           & 14h17m53.7s & 52d34m46.2s & 0.683 & 0.119 & 0.55 (4) & -    & -    & -\\
        \textbf{EGSXG\,J1420.5+5308}  & 14h20m33.3s & 53d08m21.1s & 0.734 & 0.160 & 1.07 (4) & 92   & 11   & 81 \\
        MS\,1137+6625                 & 11h40m22.8s & 66d08m14.5s & 0.782 & 0.096 & 4.70 (4) & -    & -    & -\\
        RX\,J1317+2911                & 13h17m21.8s & 29d11m17.0s & 0.805 & 0.123 & 1.73 (4) & -    & -    & -\\
        \textbf{RX\,J1716+6708}       & 17h16m38.8s & 67d08m25.8s & 0.813 & 0.082 & 5.10 (1) & -    & 2.5  & 51 \\
        \textbf{EGSXG\,J1416.2+5205}  & 14h16m16.7s & 52d05m58.2s & 0.832 & 0.120 & 1.31 (4) & -    & 2.6  & 17 \\
        \textbf{RX\,J1226+3333}       & 12h26m58.2s & 33d32m48.0s & 0.890 & 0.083 & 7.80 (3) & -    & 1.9  & 28 \\
        \textbf{CDGS54}               & 10h02m01.0s & 02d13m28.6s & 0.900 & 0.093 & 2.32 (4) & 19.2 & 3.9  & ...\\
        \textbf{CL\,J1415+3612}       & 14h15m11.2s & 36d12m04.0s & 1.030 & 0.151 & 3.44 (4) & -    & 2.9  & 7.0 \\
        RX\,J0910+5422                & 09h10m45.4s & 54d22m05.0s & 1.106 & 0.101 & 0.87 (4) & -    & -    & 2.0 \\
        RX\,J0849+4452                & 08h28m58.2s & 44d51m55.1s & 1.261 & 0.099 & 2.84 (4) & -    & -    & -\\\hline
    \end{tabular}
    \label{tab:largesample}
    \tablebib{(1) \citet{ettori04}; (2) \citet{hicks08}; (3) \citet{mantz10}; (4) \citet{pascut15}}
\end{table*}

A feedback cycle is created when supermassive black holes (SMBHs) in the cores of (proto-)cluster galaxies accrete cooling gas and accelerate relativistic jets into their environment. This re-energizes the cooling gas, and is therefore understood to play a critical role in the formation and evolution of galaxy clusters \citep[e.g.,][]{mcnamara07, fabian12, gitti12}. However, observational constraints have limited our knowledge about this feedback process mainly to the local Universe, leaving a Gyr-scale blind spot on one of the most influential epochs in which this process took place: the formation and early evolution of our present-day galaxy clusters.

As the hot and diffuse intracluster medium (ICM) permeating a cluster of galaxies cools down through the emission of X-rays, it sinks down toward the gravitational center of the cluster in the form of a ``cooling flow'' \citep{fabian94}. There, it accretes onto the central brightest cluster galaxy (BCG), which typically dominates the core of the cluster, where it is expected to lead to an extremely high star-formation rate. However, as this cooling flow also feeds the central SMBH, it creates an active galactic nucleus (AGN), which releases a significant fraction of the mass-energy of the accreting gas into the cluster environment in the form of radiation (``quasar-mode'' feedback) and two relativistic jets \citep[``radio-mode'' or mechanical feedback,][]{croton06}. This prevents the ICM from rapidly cooling down and suppresses star-formation in the central galaxies.

If the AGN accretes gas at a low fraction of its Eddington rate, it primarily provides radio-mode feedback \citep[e.g.,][]{russell13}. The two relativistic jets which inflate large radio lobes in the cluster environment, excavate regions within the ICM. This interaction re-energizes the ICM, and enables the energy output of the AGN to be measured \citep[e.g.,][]{mcnamara00, nulsen02, birzan04, rafferty06}. Based on the internal energy of the radio lobes and the work required to excavate the lobe volume against the external pressure of the ICM, the total energy required to produce the radio lobes can be determined, which in combination with an estimate of the age of the lobes can be used to measure the average energy output of the AGN during its active phase.

Due to the demanding observational requirements of performing these measurements, the high-redshift (\(z>0.6\)) regime has largely remained out of reach. The diffuse relativistic plasma of the radio lobes is primarily bright toward low radio frequencies. However, as the angular resolution of a radio interferometer scales with frequency, until recently low-frequency observations only provided a low angular resolution, making it difficult to resolve the structure of the radio lobes. Diverting to higher frequencies, where the angular resolution improves, presents the problem that the old steep-spectrum emission tends to fall below the detection limit of the instrument. Likewise, X-ray observations, which reveal ``cavities'' in the ICM coincident with the radio lobes, require infeasibly long exposure times for high-redshift clusters to reach the photon count statistics needed to identify and constrain the dimensions of the cavities.

The recent breakthrough in the calibration of LOw Frequency ARray \citep[LOFAR,][]{haarlem13} data solves the challenges presented by the strong ionospheric effects at low frequencies, the heterogeneous dipole arrays and the terabyte-scale data volumes to enable the use of the international baselines \citep{morabito21, sweijen22}. Compared to the Dutch part of LOFAR, which offers an angular resolution of \(\theta\approx6~\mathrm{arcseconds}\) at 144~MHz, the International LOFAR Telescope (ILT) is an order-of-magnitude improvement by achieving an angular resolution of \(\theta\approx0.3~\mathrm{arcseconds}\) at the same frequency. This provides the combination of angular resolution and sensitivity required to study the radio lobes in detail at high redshifts, opening the observational window for high-redshift measurements of the amount of radio-mode feedback \citep{timmerman22}.

To take advantage of this new observational frontier, we use a sample of high-resolution dedicated observations taken with the ILT to confirm the feasibility of using high-resolution low-frequency radio observations to measure the AGN energy output at high redshifts (\(z>0.6\)) for the first time. This provides both the energy output of a number of AGNs in this redshift regime for the first time as well as an initial estimate of the average success rate.

In this paper, we adopt a \(\Lambda\)CDM cosmology with a Hubble parameter of \(H_0\ =\ 70~\mathrm{km}\ \mathrm{s}^{-1}\ \mathrm{Mpc}^{-1}\), a matter density parameter of \(\Omega_m\ =\ 0.3\), and a dark energy density parameter of \(\Omega_\Lambda\ =\ 0.7\). We define our spectral indices \(\alpha\) according to \(S_\nu \propto \nu^\alpha\), where \(S_\nu\) is flux density and \(\nu\) is frequency. All uncertainties denote the 68.3\%=1\(\sigma\) confidence interval.

\begin{figure*}
    \centering
    \includegraphics[width=\textwidth]{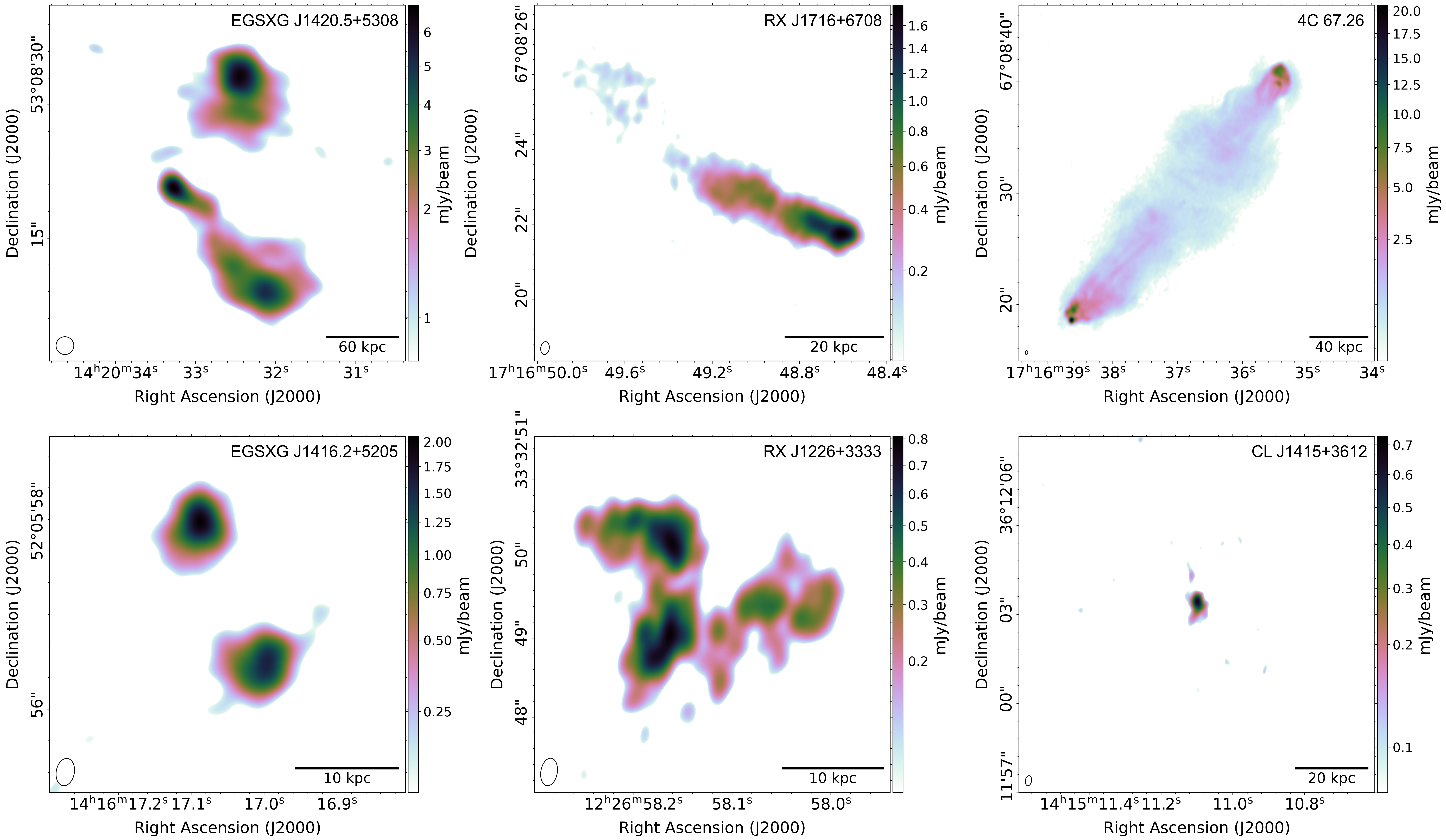}
    \caption{High-resolution LOFAR images of all detected BCGs in the galaxy clusters in our sample. The radio source 4C 67.46 was added as this galaxy of RX\,J1716+6708 is a prominent member of the cluster at radio wavelengths and may be confused with the BCG at low angular resolutions. The color maps range from three times the rms noise level to the peak brightness. The scale bar in the bottom right corner of each panel measures the listed physical length at the redshift of the respective clusters. The size of the beam is indicated by the black ellipse in the bottom left corner of each panel.}
    \label{fig:images}
\end{figure*}

\section{Methodology}

The gold standard for estimating the amount of energy injected by the central AGN into the cluster environment is computing the \textit{cavity power}. In short, as the radio jets from the AGN slow down within the cluster environment, they expand against the external pressure from the ICM. In X-ray observations, which are primarily sensitive to the hot ICM, this results in surface brightness depressions at the location of the radio lobes, also known as cavities. The total energy associated with this radio lobe is the sum of both the work required to inflate the volume of the radio lobes against the external pressure and the internal energy of the radio lobes. Assuming the radio lobes consist of relativistic gas, the total enthalpy of a radio lobe or X-ray cavity (\(E_\mathrm{cav}\)) can be calculated as
\begin{equation}
    E_\mathrm{cav} = 4pV,
\end{equation}
where \(p\) is the pressure within the ICM and \(V\) is the volume of the radio lobe. Finally, the average power output of the AGN can be obtained by dividing this total enthalpy by the age of the structures. As the radio lobes are less dense than their surrounding ICM, these can be assumed to rise away from the central AGN buoyantly. This gives an age estimate of the radio lobes of
\begin{equation}
    t_\mathrm{buoy} = R\sqrt{\frac{SC}{2gV}},
\end{equation}
where \(R\) is the distance between the AGN and the center of the radio lobe, \(S\) is the cross-sectional area of the radio lobe, \(C\) is the drag coefficient which by simulations is estimated to be around 0.75 \citep{churazov01}, and \(g\) is the local gravitational acceleration. Following \citet{birzan04}, we assume that the local acceleration is mainly the result of the mass of the BCG, and therefore we use the approximation of a isothermal sphere, such that
\begin{equation}
    g = \frac{2\sigma^2}{R},
\end{equation}
where \(\sigma\) is the stellar velocity dispersion \citep{binney01}. It would be informative to perform simulations to determine how radio lobes rise away from the AGN as a function of their environment. However, the buoyancy time scale using this approximation commonly provides an estimate in between the sound speed time scale and the refill time scale for both small and large radii \citep{birzan04, rafferty06}, making the use of this time scale preferable for now. As the stellar velocity dispersion is difficult to estimate at high redshifts based on shallow photometry, but typically does not strongly vary, we adopt an estimate of \(\sigma=289\ \mathrm{km/s}\), following \citet{birzan04}. The sample considered by \citet{birzan04} featured a scatter of 12\% on the velocity dispersion of the BCG, and therefore we assume that this standard estimate for the velocity dispersion results in an additional 12\% uncertainty on the cavity power.

The two key unknowns in this method are the dimensions of the radio lobes or cavities, and the pressure within the ICM. In the hybrid approach demonstrated in \citet{timmerman22}, high-resolution low-frequency radio observations are used to determine the dimensions and positions of the radio lobes, and X-ray observations are used to constrain the pressure within the ICM. The advantage of using radio observations to measure the size of the radio lobes rather than X-ray observations to measure the size of the cavities is that cavities are only visible as a surface brightness depression relative to the rest of the ICM. This means that it can take infeasibly long observation times for X-ray telescopes to reach the sensitivity required to detect the cavities. Additionally, cavities tend to be poorly constrained toward large radii, as the ICM also features a decreasing surface brightness profile. This is largely resolved by the ability of low-frequency radio observations to detect the old electron population of radio lobes, which typically trace the complete volume of the structure relatively well. However, currently the only instrument which has the demonstrated combination of resolution and sensitivity to the radio lobes is the ILT. By relying on radio observations to provide the volume measurements, X-ray observations only need to provide the pressure in the ICM, which requires less sensitivity.

The dimensions of the radio lobes forms the primary cause of uncertainty on the final cavity power estimates. The main uncertainty is the projection angle. To obtain the best estimate for the cavity volume, we use a Monte Carlo method to guess a random projection angle and calculate the corresponding true dimensions of the radio lobe assuming an ellipsoidal shape, the distance to the AGN core and ICM pressure. The best estimates and uncertainties are then taken as the median and 68.3\% confidence interval, respectively. Contrary to in \citet{timmerman22}, we also include the pressure profile in the Monte Carlo simulation. This was previously not possible as only the ICM pressure at the cavity position was available in literature.

We note that further uncertainties are introduced by inaccuracies in, for instance, the assumption for the pressure profile, the possibility of radio emission extending below the detection limit and the deviation of the radio lobe morphology from an ellipsoid. However, quantifying these uncertainties is not within the scope of this project.

\section{Sample}

The sample analyzed in this pilot project is based on the galaxy cluster samples of \citet{santos08}, \citet{santos10} and \citet{pascut15}. For all 88 clusters in these samples, their dynamical state was estimated based on X-ray observations. In particular, the concentration parameters of the X-ray surface brightness profile (\(c_\mathrm{SB}\)), which is estimated using the ICM column density, was measured for each of the clusters. Following \citet{santos08}, we classify non-cool core clusters as \(c_\mathrm{SB}<0.075\), weak cool-core clusters as \(0.075<=c_\mathrm{SB}<0.155\) and strong cool-core clusters as \(0.155<c_\mathrm{SB}\). As previous work, which forms the foundation of this project, is primarily focused on cool-core clusters, we disregard non-cool core clusters for high-resolution observations in this pilot project. However, we do intend to explore the non-cool core regime in a follow-up project. Finally, we can only consider galaxy clusters located in the Northern Hemisphere due to the observational constraints of LOFAR. After selecting only cool-core (\(c_\mathrm{SB}>0.075\)) galaxy clusters in the Northern hemisphere at high-redshift (\(z>0.6\)), a sample of thirteen galaxy clusters remains, which forms our set of targets of interest. The basic properties of these objects are summarized in Table \ref{tab:largesample}.

To derive the cavity power associated with the central AGNs of these clusters, high-resolution LOFAR observations have been carried out. As processing such observations is computationally expensive, we only proceeded to analyze galaxy clusters which are sufficiently bright (>5~mJy at 144~MHz) in the LOFAR Two-Metre Sky Survey \citep[LoTSS,][]{shimwell17, shimwell19, shimwell22} to potentially provide a significant result. Sources outside of the current LoTSS coverage were also included if the flux densities observed in the TIFR Giant Metrewave Radio Telescope Sky Survey \citep[TGSS,][]{intema17} and the Very Large Array Sky Survey \citep[VLASS,][]{lacy20} suggest that they would likely exceed 5~mJy in brightness at 144~MHz. This final subsample consists of EGSXG\,J1420.5+5308, RX\,J1716+6708, EGSXG\,J1416.2+5205, RX\,J1226+3333, CCDGS54 and CL\,J1415+3612. The remaining targets are considered non-detections for standard ILT observations.

\begin{figure*}
    \centering
    \includegraphics[width=\textwidth]{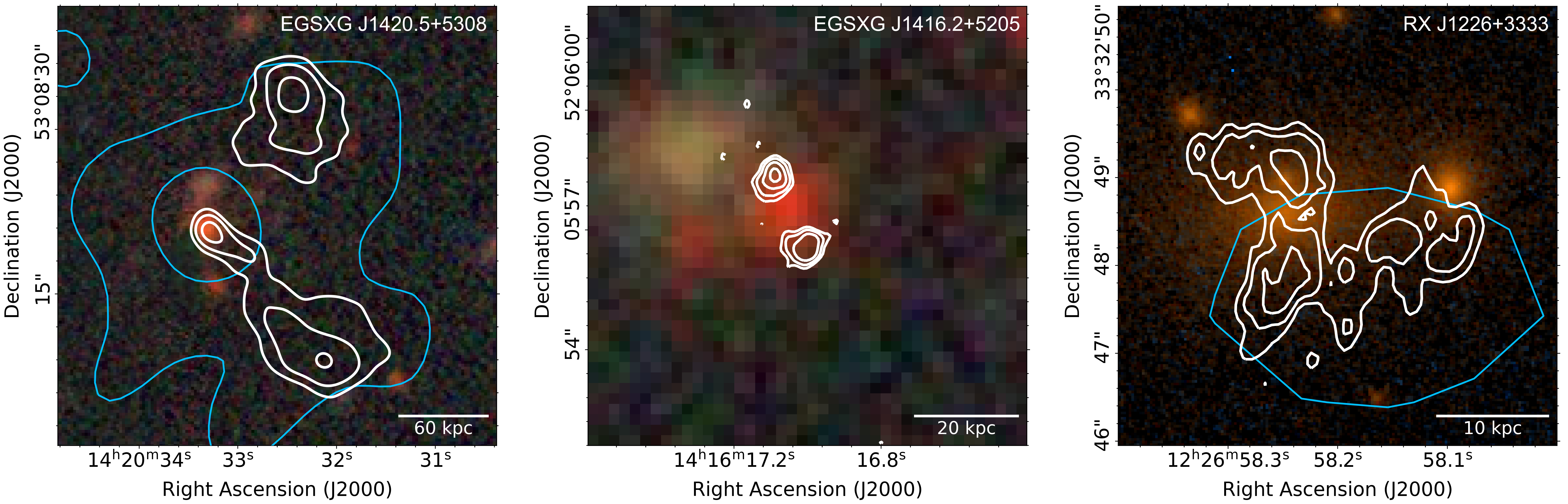}
    \caption{Optical images of the three BCGs in our sample for which radio lobes have been detected with LOFAR. The optical images for the two EXSXG clusters were obtained from the DESI Legacy Imaging Survey \textit{g}, \textit{r} and \textit{z} bands \citep{dey19}, whereas for RX\,J1226+3333 the optical image was obtained from Hubble Space Telescope observations with the ACS/WFC detector in F850LP, F625W and F435W filters. The white contours indicate the radio emission at 144~MHz and are drawn at \(4\sigma_\mathrm{rms}\) and increase in factors of 2. The blue contours indicate the 0.5--7~keV X-ray emission as detected by the Chandra X-Ray Observatory. The scale bar in the bottom right corner of each panel measures the listed length at the redshift of the respective cluster.}
    \label{fig:overlays}
\end{figure*}

\section{Observations and data reduction}

\subsection{Radio - LOFAR}

The sources selected for high-resolution observations were observed with LOFAR for a duration of 8 hours each in the frequency range between 120 and 168~MHz. A summary of these observations is provided in Table~\ref{tab:LOFAR}. To be able to perform the clock and bandpass calibration, a calibrator source was observed for 10 minutes before and after the target observation. As a first step, the data were reduced using the standard LOFAR calibration \textsc{Prefactor} software package \citep{vanweeren16, williams16, gasperin19}. This step includes flagging radio-frequency interference (RFI) using \textsc{AOFlagger} \citep{offringa13, offringa15}. Next, a model of the calibrator source was compared to the data to calculate corrections for the delays between the two polarizations per station, the Faraday rotation due to magnetic fields in Earth's ionosphere, the bandpass and the clock offsets between the different LOFAR stations. With these corrections applied to the target data, RFI was again flagged from these data before these are averaged to a time resolution of 8~seconds per integration and 98~kHz of bandwidth per frequency channel. Finally, the data from the Dutch part of LOFAR were compared to a TGSS sky model to perform the initial phase-only calibration of these visibilities.

Following the standard LOFAR calibration pipelines, we proceeded with the LOFAR-VLBI pipeline developed by \citet{morabito21}. This pipeline starts by applying the previously derived calibration solutions to all LOFAR stations. Next, LOFAR's core stations were phased up to form a single large virtual station with a narrow field of view, thereby reducing the interference from unrelated nearby radio sources on the target. Then, dispersive phase corrections and residual gain corrections were derived by self-calibrating on a bright and compact source from the Long-Baseline Calibrator Survey \citep[LBCS,][]{jackson16, jackson21} near the target. Finally, after all calibration solutions were applied to the target, a final self-calibration routine of total electron content (TEC) and phase fitting using \textsc{DP3} mode 'tecandphase' was performed on the target source. This accounts for the direction dependence of the previously derived calibration solutions \citep{vanweeren21}.

\begin{table*}
    \caption{Summary of the LOFAR observations processed for this paper. For each cluster, the project code of the original proposal is listed along with with the PI of the proposal and the date and duration of the corresponding observations.}
    \centering\small
    \begin{tabular}{lllll}\hline\hline
        Cluster name & Project code & PI & Date & Duration\\\hline
        EGSXG\,J1420.5+5308 & LC16\_015 & Timmerman & 16 July 2021 & 8 hours \\
        RX\,J1716+6708 & LC11\_016 & Bempong-Manful & 23 December 2018 & 8 hours \\
        EGSXG\,J1416.2+5205 & LC16\_015 & Timmerman & 16 July 2021 & 8 hours \\
        CDGS54 & LC15\_031 & Timmerman & 11 December 2020 & 2 hours \\
        & LC15\_031 & Timmerman & 18 December 2020 & 2 hours \\
        & LC15\_031 & Timmerman & 21 December 2020 & 2 hours \\
        & LC15\_031 & Timmerman & 25 December 2020 & 2 hours \\
        CL\,J1415+3612 & LC16\_015 & Timmerman & 25 Nov 2021 & 8 hours \\\hline\hline
    \end{tabular}
    \label{tab:LOFAR}
\end{table*}

\subsection{X-rays - Chandra}

We queried the Chandra data archive\footnote{\url{https://cda.harvard.edu/chaser/}} and downloaded the Chandra Advanced CCD Imaging Spectrometer (ACIS) observations of the targets in our sample that show clear radio lobes (Fig.~\ref{fig:images} and Table~\ref{tab:cavities}) to obtain X-ray images of the thermal gas emission. We retrieved data for EGSXG J1420.5+5308, EGSXG J1416.2+5205 and RX\,J1226+3333, as summarized in Table~\ref{tab:xray-obs}. All data were reduced using CIAO v4.12 tools following standard procedures, adopting the CALDB v4.9.0 (see also our previous work, \citealt{timmerman22}). Multiple ObsIDs for the same target were mosaiced into a single image in the 0.5-7.0 keV band for the subsequent analysis.

\begin{table}[]
    \caption{Summary of Chandra X-ray observations processed for this paper. For each cluster, the relevant Observation IDs, the detector with which these observations were taken, and the total net exposure time are listed. All observations were carried out in VFAINT mode.}
    \centering
    \begin{tabular}{llll}\hline\hline
    Cluster name & Net time & ObsIDs & Detector \\
                 & (ks)              &        &          \\\hline
    EGSXG\,J1420.5+5308 & 681.7 & 5845-5846 & ACIS-I \\
                       &     & 6214-6215 & ACIS-I \\
                       &     & 9450-9452 & ACIS-I \\
                       &     & 9720-9726 & ACIS-I \\
                       &     & 9793-9797 & ACIS-I \\
                       &     & 9842-9844 & ACIS-I \\
                       &     & 9863 & ACIS-I \\
                       &     & 9866 & ACIS-I \\
                       &     & 9870 & ACIS-I \\
                       &     & 9873 & ACIS-I \\
                       &     & 9876 & ACIS-I \\
    EGSXG\,J1416.2+5205 & 179.3 & 5853-5854 & ACIS-I \\
                       &     & 6222-6223 & ACIS-I \\
                       &     & 6366 & ACIS-I \\
    RX\,J1226+3333     & 69.9 & 932 & ACIS-S \\
                       &     & 3180 & ACIS-I \\
                       &     & 5014 & ACIS-I \\\hline\hline

    \end{tabular}
    \label{tab:xray-obs}
\end{table}

\begin{table*}
\caption{Properties of the cavities in our sample derived using the hybrid X-ray--radio method.}
    \renewcommand{\arraystretch}{1.2}
    \centering\small
    \begin{tabular}{lllllllll}\hline\hline
        (a)          & (b)      & (c)            & (d)            & (e)   & (f) & (g) & (h)               & (i)\\
        Cluster name & Redshift & \(R_\text{l}\) & \(R_\text{w}\) & \(R\) & V & \(p\) & \(t_\text{buoy}\) & \(P_\text{cav}\)\\
        & & (kpc) & (kpc) & (kpc) & (kpc\(^3\)) & (keV/cm\(^3\)) & (10\(^7\) yr) & (10\(^{42}\) erg/s)\\\hline
        EGSXG\,J1420.5+5308 (N) & 0.734 & 43.3 & 40.0 & 105  & \(3.04^{+0.35}_{-0.21} \times 10^5\) & \(6.15^{+3.05}_{-3.02} \times 10^{-3}\) & \(25.5^{+23.1}_{-4.7}\) & \(42.9^{+27.3}_{-28.4}\) \\
        EGSXG\,J1420.5+5308 (S) & 0.734 & 51.7 & 35.7 & 112  & \(3.03^{+1.14}_{-0.29} \times 10^5\) & \(5.77^{+2.99}_{-2.96} \times 10^{-3}\) & \(25.2^{+17.9}_{-4.2}\) & \(41.1^{+24.0}_{-24.1}\) \\
        EGSXG\,J1416.2+5205 (N) & 0.832 & 3.04 & 2.66 & 7.53 & \(93.8^{+55.5}_{-38.9}\) & \(1.86^{+1.20}_{-0.75} \times 10^{-2}\) & \(1.36^{+1.38}_{-0.56}\) & \(0.49^{+0.64}_{-0.29}\) \\
        EGSXG\,J1416.2+5205 (S) & 0.832 & 2.28 & 2.55 & 7.53 & \(57.9^{+38.4}_{-26.6}\) & \(1.87^{+1.20}_{-0.75} \times 10^{-2}\) & \(2.19^{+1.42}_{-0.65}\) & \(0.28^{+0.40}_{-0.18}\) \\
        RX\,J1226+3333 (N)      & 0.890 & 5.47 & 2.64 & 6.21 & \(195^{+132}_{-82}\) & \(1.03^{+1.75}_{-0.63}\) & \(1.06^{+0.49}_{-0.24}\) & \(108^{+233}_{-73}\) \\
        RX\,J1226+3333 (S)      & 0.890 & 5.32 & 1.79 & 6.21 & \(89.4^{+84.7}_{-50.1}\) & \(1.03^{+1.75}_{-0.62}\) & \(1.07^{0.67}_{-0.33}\) & \(46.7^{+114.6}_{-34.8}\) \\\hline
    \end{tabular}
    \tablefoot{
        Columns: (a) cluster name; (b) cluster redshift; (c) cavity radius along the jet axis; (d) cavity radius perpendicular to the jet axis; (e) central cavity distance from the AGN core; (f) cavity volume; (g) ICM pressure at the projected distance of the radio lobe to the AGN core; (h) buoyancy timescale; (i) cavity power
    }
    \label{tab:cavities}
\end{table*}

\section{Results}

\subsection{Imaging}

To derive the cavity power associated with the AGNs in our high-redshift galaxy cluster sample, we produced high-resolution (\({\sim}0.3\)~arcseconds) images at 144~MHz using the calibrated data sets of all sources in our sample using \textsc{WSClean} \citep{offringa14, offringa17}. Of these images, only CDGS54 resulted in a non-detection, suggesting that this source contains mostly emission on angular scales significantly larger than 0.3~arcseconds. The other galaxy clusters are shown in Fig.~\ref{fig:images}. We included the radio source 4C\,67.26 in this figure as it was identified to be a member galaxy of RX\,J1716+6708 \citep{gioia99} and it dominates the region at radio wavelengths. 

Our images reveal clear sets of diffuse radio structures opposite and equidistant from their host galaxies in EGSXG\,J1420.5+5308 and 4C\,67.26, which we therefore identify as radio lobes associated with the AGN. As shown in Fig.~\ref{fig:overlays}, the radio lobes in EGSXG\,J1420.5+5308 originate from the BCG in the cluster, which in turn lies at the central ICM peak. We detect two additional sets of radio lobes in EGSXG\,J1416.2+5205 and RX\,J1226+3333, but it is unclear whether we have revealed the full physical extent of these radio lobes as these lobes are fainter, more compact, and therefore fail to provide a clear indication for the presence of a sharp boundary associated with the physical limit of the structure. The radio source 4C\,67.26 shows a strong and undisturbed FR II-like morphology, but as this member galaxy is located far outside of the cluster center and therefore not part of the feedback cycle, it will be disregarded from further analysis.

The radio structure of EGSXG\,J1416.2+5205 consists of two equally bright components, which are assumed to be the radio lobes of the BCG, but due to the barely resolved structure, this remains uncertain. As shown in Fig.~\ref{fig:overlays}, the BCG associated with EGSXG\,J1416.2+5205 is located directly in between the two radio lobes. Similarly, RX\,J1226+3333 also features uncertain radio lobes. Given that the BCG is detected directly in between the eastern-most structures, these are considered to be two radio lobes located toward the North and South of the BCG, with the western-most emission likely being unrelated. However, due to the low brightness of the radio lobes, it is unclear if the full extent of these structures is detected. The peak in X-ray surface brightness near the BCG of RX\,J1226+3333 confirms that its AGN outflows lie at the core of the cluster.

The BCG of RX\,J1716+6708 shows radio-emitting plasma bending toward the East. This aligns with the direction of the distribution of members galaxies \citep{henry97}, suggesting that the galaxy cluster may be in the late stage of a merger event, consistent with its relatively low concentration parameter of \(c_\mathrm{SB}=0.082\). Finally, CL\,J1415+3612 shows a single compact radio component coincident with the BCG in the core of the cluster. Unfortunately, due to the low count statistics of the associated X-ray maps, the identification of radio lobes can not be aided by X-ray observations for our sample.

\subsection{Analysis}

From our radio images, we estimate the volume of the radio lobes we have detected assuming an ellipsoidal shape. We note that we consider our observation of EGSXG\,J1420.5+5308 to allow a reliable radio lobe volume measurements, and consider these measurements to be tentative for EGSXG\,J1416.2+5205 and RX\,J1226+3333. 

As our sample is selected to be at high redshifts, the X-ray observations do not contain sufficient photon counts to obtain pressure profiles for the individual clusters. Therefore, we use the average pressure profile of a sample of galaxy clusters at high redshift (\(0.6<z<1.2\)) from \citet{mcdonald14}, which is normalized by $P_{500}$ and $r_{500}$, to estimate the ICM thermal pressure at the distance where we observe the radio lobes in our clusters. For EGSXG\,J1420.5+5308 and EGSXG\,J1416.2+5205, we use the $M_{500}$ reported in Table~\ref{tab:largesample} to compute the $P_{500}$ expected in the standard self-similar model \citep[e.g.,][]{nagai07}. For RX\,1226+3333, instead the pressure profile of the ICM was determined by \citet{romero18} using Sunyaev-Zel'dovich (SZ) observations and therefore we directly adopt the value measured via SZ. For the pressure acting on the radio lobe, the ICM pressure at the central position of the radio lobes is used, as this has been found to be a reasonable estimate for the pressure surrounding the entire radio lobe \citep{hardcastle13}.

Using the measured volumes of the radio lobes, the ICM pressure estimates and our assumption for the stellar velocity dispersion of the BCGs, we calculate the final cavity power as the ratio between the cavity enthalpy and the buoyancy timescale. Our measurements for the radio lobe dimensions and local ICM pressures are summarized together with the resulting cavity powers in Table~\ref{tab:cavities}. We find that EGSXG\,J1420.5+5308 and RX\,J1226+3333 contain the most radio-mode feedback of our sample, with cavity powers of around \(84.3\times10^{42}\) erg/s and \(155\times10^{42}\) erg/s, respectively. Meanwhile, the small radio lobes detected in EGSXG\,J1416.2+5205 only correspond to a cavity power of around \(0.77\times10^{42}\) erg/s in total.

\section{Discussion}

The operation of radio-mode feedback has remained poorly quantified at high redshifts (\(z>0.6\)). Although significant progress has been made using X-ray observations \citep[e.g.,][]{hlavacek15}, the observational requirement to clearly detect cavities in the ICM at high redshifts forms a bottleneck. In \citet{timmerman22}, we verified at low redshifts that high-resolution low-frequency radio observations taken with the ILT can be used in conjunction with X-ray observations to measure the output power of the AGN, thereby making these measurements more feasible. In this paper, we have applied this method for the first time on a high-redshift sample of galaxy clusters to obtain the first measurements of the energy injected by the AGN into its cluster environment for these systems and confirm the feasibility on this hybrid method in this regime.

\subsection{Radio lobe detections}

Of the original high-redshift sample of 13 cool-core galaxy clusters compiled by \citet{santos08}, \citet{santos10} and \citet{pascut15}, 6 had previously been detected at radio wavelengths and were therefore selected for high-resolution imaging. Of these 6 clusters, we detected radio lobes associated with 3 BCGs and one additional cluster member. This corresponds to a 23\% detection rate of radio lobes, which is significantly higher than the 4.7\% detection rate (2 out of 43) previously achieved using X-ray observations of the SPT-SZ sample within the same redshift range \citep{mcdonald13,bleem15,hlavacek15,bocquet19}. However, we note that two of our systems were not bright enough to confirm whether or not the radio lobes extend below the sensitivity limit of our observations. Therefore, the radio lobe volumes and therefore also the associated cavity powers can conservatively be considered to form lower limits on the true cavity volume and hence power.

In \citet{timmerman22}, we stated that radio lobes could be expected to be reliably detected in systems brighter than approximately \(\sim100\)~mJy at 144~MHz. Our clear detection of radio lobes in EGSXG\,J1420.5+5308 with a total flux density of 81~mJy is consistent with this rough previous estimate. Additionally, our radio lobe detections in EGSXG\,J1416.2+5205 with 17 mJy and RX\,J1226+3333 with 28~mJy demonstrate that it is possible to detect compact radio lobes at lower flux densities. However, a lack of radio lobe detections in RX\,J1226+3333, CL\,J1415+3612 and CDGS54 also demonstrate the sensitivity limitation for 8-hour ILT observation, even for relaxed clusters (e.g., \(c_\mathrm{SB}=0.151\) for CL\,J1415+3612) and decent flux densities (e.g., 51~mJy for RX\,J1226+3333).

To probe the range of cavity powers accessible to the angular resolution of ILT observations at high redshifts, assuming the sensitivity of an 8-hour observation, we calculate the approximate smallest radio lobe which the ILT can resolve. This defines the lowest cavity power to which the ILT is sensitive. We assume a radio lobe size similar to the angular size of the synthesized beam of the ILT, a distance between the radio lobe and the AGN of two beams to provide a clear identification as a radio lobe, and a pressure profile corresponding to a low-mass galaxy cluster of \(10^{14}\ \mathrm{M}_\odot\). For comparison, we perform the same calculation for the standard 6-arcsecond angular resolution observations of only the Dutch part of the ILT. Both exclusion regions are plotted in Fig. \ref{fig:Pcav_z}. This shows that the ILT at high redshifts is sensitive to cavity powers on the order of \(10^{42}\) erg/s or more, which enables it to reach even the low-power AGNs in the cores of high-redshift galaxy clusters.

\subsection{Balance of heating and cooling}

As one of the main motivations for studying the energy output of an AGN in a cluster is its role in the feedback cycle with the ICM, it is useful to compare the derived cavity power to the X-ray luminosity in the core region of the cluster. Due to the very low count statistics for the other two targets, we can only derive an X-ray luminosity for the core of RX\,J1226+3333. Adopting the \(R_{500}\) value from \citet{mantz10} which is also adopted by \citet{romero18} for the SZ-derived ICM pressure profile, and the average \(R_{500}/R_\mathrm{cool}\) ratio determined by \citet{mittal11}, we find a bolometric X-ray luminosity within \(R_\mathrm{cool}\) of \(6.6\times10^{44}\ \mathrm{erg/s}\). Because of the low count statistics, we were not able to derive a deprojected estimate. For EGSXG\,J1416.2+5205, we adopt the X-ray luminosity determined by \citet{erfanianfar13}, who report a value of \(L_X = 46.9\times10^{42}\ \mathrm{erg/s}\) in the 0.1--2.4~keV band. We note that the bolometric luminosity will be at least higher than this estimate. Finally, for EGSXG\,J1420.5+5308, we adopt the estimate reported by \citet{jeltema09}, who find a bolometric X-ray luminosity within 364~kpc of \(37.3\times10^{42}\ \mathrm{erg/s}\). As we estimate that the cooling radius will be only 38.5~kpc, this should be taken as an upper limit on the X-ray luminosity within the cooling radius.

Comparing these X-ray luminosities to our cavity power measurements, we should distinguish between our reliable and unreliable radio lobe detections. With EGSXG\,J1420.5+5308, we find that the cavity power exceeds the X-ray luminosity by a factor of approximately 2.3. This is consistent with the average of the population, as demonstrated by \citet{gitti12}. For low-mass systems, they find that the heating typically exceeds the cooling by over a factor of 5, whereas for high-mass systems, this decreases to only 1.51. Given the intermediate cavity power and cooling rate of EGSXG\,J1420.5+5308, an excess of a factor 2.3 is consistent with lower-redshift results. In EGSXG\,J1416.2+5205, where the detection of the radio lobes was marginal, we find that the cooling strongly exceeds the heating by a factor of over 60. This mostly reaffirms that the radio lobe detections should be considered tentative, though we note that the X-ray luminosity is difficult to obtain with the very low count statistics as well. Finally, in RX\,J1226+3333, we obtain a cooling excess of approximately a factor of 4.3, which is closer to the general population as reported by \citet{gitti12}. As these radio lobes are also not clearly detected, we defer any interpretations of these ratios until after more or deeper observations have been used to determine the cavity power of high-redshift galaxy clusters.

\subsection{Pressure profiles}

Measuring the pressure profiles on a per-cluster basis is impossible due to the low-count statistics of the available X-ray observations. To obtain the ICM pressure at the position of the radio lobes, we used the average pressure profiles from \citet{mcdonald14} derived for a sample of high-redshift clusters. Although this enabled us to determine the cavity power of the AGNs in our sample using literature results derived from shallower X-ray observations compared to those required for determining the cavity volume, we note that obtaining the ICM pressure is still affected by significant uncertainties. In addition to the intrinsic scatter among pressure profiles, the standard pressure profiles from \citet{mcdonald14} depend on \(M_{500}\), which in turn is typically derived based on scaling relations such as the temperature--mass relation \citep[e.g.,][]{sun09}. While X-ray observations would still be required to establish the mass and dynamical state of the clusters, SZ observations can also be employed to directly provide pressure profiles of the ICM. As this circumvents the aforementioned systematic uncertainties, we suggest that a combination of X-ray and high-resolution SZ observations with instruments such as MUSTANG and NIKA should be employed at higher redshifts where possible.

\subsection{Comparison to previous detections}

\begin{figure}
    \centering
    \includegraphics[width=\columnwidth]{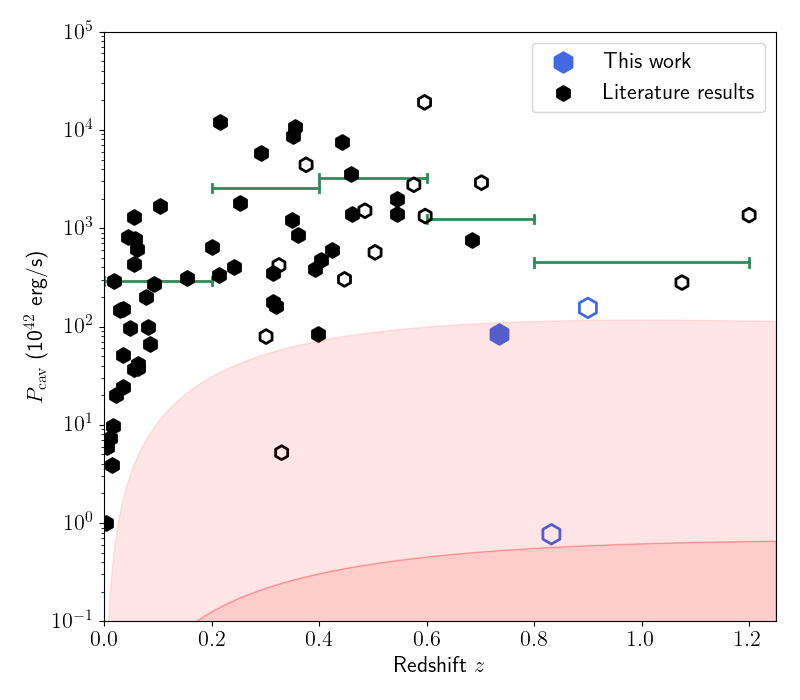}
    \caption{Cavity power per galaxy cluster as a function of redshift. The blue data points indicate the three radio lobe systems detected in our sample, and the black data points indicate the cavity power estimates published by \citet{rafferty06}, \citet{hlavacek12}, \citet{hlavacek15} and \citet{timmerman22}. Cavity/radio lobe systems which were indicated to be low significance are indicated with open data points, whereas reliable detections are indicated with solid data points. The green bars show the average cavity power within each redshift bin, indicated by the width of each bar. The light red region indicates the approximate region which is not accessible by 6-arcsecond observations, such as those taken with only the Dutch part of LOFAR. The darker red region indicates the approximate region which is not accessible by 0.3-arcsecond observations, such at those taken with the complete ILT.}
    \label{fig:Pcav_z}
\end{figure}

Cavity powers within our redshift regime have only previously been published by \citet{hlavacek15}, who used deep X-ray observations of the SPT-SZ sample. They detected clear cavities at \(z=0.6838\) and potential cavities at \(z=0.7019\). Additionally, they report marginally convincing cavities at \(z=1.075\) and \(z=1.2\). This is largely in agreement with our radio lobe detections at \(z=0.734\), \(z=0.823\) and \(z=0.890\), which simultaneously almost double the amount of measurements at \(z>0.6\) and establish a new highest redshift at which the cavity power is constrained using radio observations. To compare our cavity power measurements with the results from \citet{hlavacek15}, we have plotted these measurements together with additional low-redshift measurements in Fig.~\ref{fig:Pcav_z}. Our measurements are largely in line with the general trend of cavity power as a function of redshift as probed by \citet{rafferty06}, \citet{hlavacek12} and \citet{hlavacek15}. The current data, though still consistent with a constant cavity power level, appear to suggest that the average cavity power peaked at a redshift of \(z\sim0.4\) and slowly decreases toward higher redshifts. However, the scatter within each bin is relatively large, and the higher-redshift bins are strongly affected by low-significance cavity power measurements. Therefore, we both require more and tighter constraints on the cavity volume and a better understanding of our observational systematics to confirm any deviation of the cavity power trend from a constant level.

\subsection{Future projection}

As we have demonstrated in \citet{timmerman22} and this pilot project, it is feasible to investigate samples of galaxy clusters with the ILT. For the purpose of this pilot project, we have only considered cool-core galaxy clusters. However, this simultaneously limited the sample to previously well-studied clusters and excluded a significant fraction of all clusters from this analysis. It would be a logical next step to investigate a wider selection of galaxy clusters at high redshifts. How this would affect the average cavity power trend compared to for only cool-core clusters currently remains an open question. The exact ratio between cool-core clusters and non-cool-core clusters as a function of redshift strongly depends on the adopted metric \citep{mcdonald13}, illustrating that the differences between these two categories are poorly understood at high redshifts. Larger catalogs of galaxy clusters such as the catalogs compiled by \citet{wen18}, \citet{wen21} and \citet{wen22} provide a valuable sample of optically-selected high-redshift galaxy clusters in the Northern hemisphere, which can be observed by the ILT. Additionally, the recently launched Euclid mission is projected to detect more than \(10^5\) galaxy clusters up to a redshift of \(z=2\) \citep{euclid19}. By following up cluster detections from these surveys to perform cavity power measurements, the trend of radio-mode feedback across cosmic time can be better understood for the complete population of galaxy clusters. It should be noted that it also remains uncertain to what extent these measurements can reliably be performed for non-cool-core clusters. Finally, future SZ surveys in the Northern hemisphere would be critical to ensure that we also investigate the low cluster mass regime at high redshifts.

It is likely that a larger systematic survey containing at least a few dozen cavity power measurements will provide a statistically reliable indication of the trend of radio-mode feedback up to \(z\sim 1.2\). Given the current success rate, this will probably require on the order of a hundred high-resolution ILT maps of galaxy clusters in this redshift regime. The current LoTSS DR2 \citep{shimwell22} provides ILT coverage of a large fraction of the Northern sky, meaning that the raw observations to perform this imaging are readily available from the LOFAR archive, leaving only the processing, calibration and imaging. While still computationally demanding (a single postage stamp ILT image requires \(\sim50,000\) core hours), ongoing progress with the development of the LOFAR-VLBI pipelines \citep{morabito21} is bringing these sample sizes within reach.  In addition to the observations, it should also be more carefully considered how non-detections should be interpreted, as well as the statistical completeness in terms of primarily mass and dynamical state, both at low and high redshifts.

\section{Conclusions}

In this paper, we have applied the hybrid method of measuring cavity powers \citep{timmerman22} using combined radio and X-ray observations to a high-redshift (\(z>0.6\)) sample of cool-core galaxy clusters. Out of the 13 galaxy clusters within our sample, we were able to detect radio lobes associated with the BCG of three clusters and resolve these using LOFAR's international baselines. Using the pressure profiles by \citet{mcdonald14}, we estimated the ICM pressure surrounding the radio lobes. Combined, these measurements provided the first cavity power estimates of three high-redshift galaxy clusters, adding to a very limited sample of these measurements at high redshifts. This demonstrates that the hybrid method of measuring cavity powers is indeed viable at high redshifts using the ILT, which encourages further work in this epoch.

Our cavity power measurements of radio-mode feedback at \(z>0.6\), do fall on the relatively low side compared to the population at lower redshifts. However, not enough measurements are currently present to provide meaningful statistics, which encourages further investigation of larger samples.

The combination of statistically representative galaxy cluster samples at high redshift, such as will be obtained by Euclid, and subsequent imaging using the ILT, will enable detailed studies of the evolution of radio-mode feedback across cosmic time. Also, such studies will improve our understanding of the differences between radio-mode feedback in cool-core clusters versus non-cool-core clusters.

\begin{acknowledgements}
    We would like to thank the anonymous referee for useful comments.
    R.T. and R.J.v.W. acknowledge support from the ERC Starting Grant ClusterWeb 804208. A.B. acknowledges financial support from the European Union - Next Generation EU. L.K.M. is grateful for support from the Medical Research Council [MR/T042842/1].
    This work made use of the Dutch national e-infrastructure with the support of the SURF Cooperative using grant no. EINF-1287. This work is co-funded by the EGI-ACE project (Horizon 2020) under Grant number 101017567.
    This paper is based (in part) on data obtained with the International LOFAR Telescope (ILT) under project codes LC15\_031 and LC16\_015. LOFAR \citep{haarlem13} is the Low Frequency Array designed and constructed by ASTRON. It has observing, data processing, and data storage facilities in several countries, that are owned by various parties (each with their own funding sources), and that are collectively operated by the ILT foundation under a joint scientific policy. The ILT resources have benefitted from the following recent major funding sources: CNRS-INSU, Observatoire de Paris and Université d'Orléans, France; BMBF, MIWF-NRW, MPG, Germany; Science Foundation Ireland (SFI), Department of Business, Enterprise and Innovation (DBEI), Ireland; NWO, The Netherlands; The Science and Technology Facilities Council, UK; Ministry of Science and Higher Education, Poland. This work made use of the Dutch national e-infrastructure with the support of the SURF Cooperative using grant no. EINF-1287. This project has received support from SURF and EGI-ACE. EGI-ACE receives funding from the European Union’s Horizon 2020 research and innovation programme under grant agreement No. 101017567. The Jülich LOFAR Long Term Archive and the German LOFAR network are both coordinated and operated by the Jülich Supercomputing Centre (JSC), and computing resources on the supercomputer JUWELS at JSC were provided by the Gauss Centre for Supercomputing e.V. (grant CHTB00) through the John von Neumann Institute for Computing (NIC). 
    This research has made use of data obtained from the Chandra Data Archive and the Chandra Source Catalog, and software provided by the Chandra X-ray Center (CXC) in the application packages CIAO and Sherpa.
    This research is based on observations made with the NASA/ESA Hubble Space Telescope obtained from the Space Telescope Science Institute, which is operated by the Association of Universities for Research in Astronomy, Inc., under NASA contract NAS 5–26555. These observations are associated with program 12791.
    The Legacy Surveys consist of three individual and complementary projects: the Dark Energy Camera Legacy Survey (DECaLS; Proposal ID \#2014B-0404; PIs: David Schlegel and Arjun Dey), the Beijing-Arizona Sky Survey (BASS; NOAO Prop. ID \#2015A-0801; PIs: Zhou Xu and Xiaohui Fan), and the Mayall z-band Legacy Survey (MzLS; Prop. ID \#2016A-0453; PI: Arjun Dey). DECaLS, BASS and MzLS together include data obtained, respectively, at the Blanco telescope, Cerro Tololo Inter-American Observatory, NSF’s NOIRLab; the Bok telescope, Steward Observatory, University of Arizona; and the Mayall telescope, Kitt Peak National Observatory, NOIRLab. Pipeline processing and analyses of the data were supported by NOIRLab and the Lawrence Berkeley National Laboratory (LBNL). The Legacy Surveys project is honored to be permitted to conduct astronomical research on Iolkam Du’ag (Kitt Peak), a mountain with particular significance to the Tohono O’odham Nation.
    NOIRLab is operated by the Association of Universities for Research in Astronomy (AURA) under a cooperative agreement with the National Science Foundation. LBNL is managed by the Regents of the University of California under contract to the U.S. Department of Energy.
    This project used data obtained with the Dark Energy Camera (DECam), which was constructed by the Dark Energy Survey (DES) collaboration. Funding for the DES Projects has been provided by the U.S. Department of Energy, the U.S. National Science Foundation, the Ministry of Science and Education of Spain, the Science and Technology Facilities Council of the United Kingdom, the Higher Education Funding Council for England, the National Center for Supercomputing Applications at the University of Illinois at Urbana-Champaign, the Kavli Institute of Cosmological Physics at the University of Chicago, Center for Cosmology and Astro-Particle Physics at the Ohio State University, the Mitchell Institute for Fundamental Physics and Astronomy at Texas A\&M University, Financiadora de Estudos e Projetos, Fundacao Carlos Chagas Filho de Amparo, Financiadora de Estudos e Projetos, Fundacao Carlos Chagas Filho de Amparo a Pesquisa do Estado do Rio de Janeiro, Conselho Nacional de Desenvolvimento Cientifico e Tecnologico and the Ministerio da Ciencia, Tecnologia e Inovacao, the Deutsche Forschungsgemeinschaft and the Collaborating Institutions in the Dark Energy Survey. The Collaborating Institutions are Argonne National Laboratory, the University of California at Santa Cruz, the University of Cambridge, Centro de Investigaciones Energeticas, Medioambientales y Tecnologicas-Madrid, the University of Chicago, University College London, the DES-Brazil Consortium, the University of Edinburgh, the Eidgenossische Technische Hochschule (ETH) Zurich, Fermi National Accelerator Laboratory, the University of Illinois at Urbana-Champaign, the Institut de Ciencies de l’Espai (IEEC/CSIC), the Institut de Fisica d’Altes Energies, Lawrence Berkeley National Laboratory, the Ludwig Maximilians Universitat Munchen and the associated Excellence Cluster Universe, the University of Michigan, NSF’s NOIRLab, the University of Nottingham, the Ohio State University, the University of Pennsylvania, the University of Portsmouth, SLAC National Accelerator Laboratory, Stanford University, the University of Sussex, and Texas A\&M University.
    BASS is a key project of the Telescope Access Program (TAP), which has been funded by the National Astronomical Observatories of China, the Chinese Academy of Sciences (the Strategic Priority Research Program “The Emergence of Cosmological Structures” Grant \# XDB09000000), and the Special Fund for Astronomy from the Ministry of Finance. The BASS is also supported by the External Cooperation Program of Chinese Academy of Sciences (Grant \# 114A11KYSB20160057), and Chinese National Natural Science Foundation (Grant \# 12120101003, \# 11433005).
    The Legacy Survey team makes use of data products from the Near-Earth Object Wide-field Infrared Survey Explorer (NEOWISE), which is a project of the Jet Propulsion Laboratory/California Institute of Technology. NEOWISE is funded by the National Aeronautics and Space Administration.
    The Legacy Surveys imaging of the DESI footprint is supported by the Director, Office of Science, Office of High Energy Physics of the U.S. Department of Energy under Contract No. DE-AC02-05CH1123, by the National Energy Research Scientific Computing Center, a DOE Office of Science User Facility under the same contract; and by the U.S. National Science Foundation, Division of Astronomical Sciences under Contract No. AST-0950945 to NOAO.
\end{acknowledgements}

\bibliographystyle{aa}
\bibliography{refs}

\end{document}